\documentclass[%
 aip,
 rsi,
 amsmath,amssymb,
reprint,%
]{revtex4-1}

\usepackage{graphicx}
\usepackage{dcolumn}
\usepackage{bm}

\usepackage[utf8]{inputenc}
\usepackage[T1]{fontenc}
\usepackage{mathptmx}
\usepackage{etoolbox}
\usepackage{pythonhighlight}
\usepackage{caption}
\usepackage{multirow}
\usepackage{array}

\graphicspath{{Figures/}}

\makeatletter
\def\@email#1#2{%
 \endgroup
 \patchcmd{\titleblock@produce}
  {\frontmatter@RRAPformat}
  {\frontmatter@RRAPformat{\produce@RRAP{*#1\href{mailto:#2}{#2}}}\frontmatter@RRAPformat}
  {}{}
}%
\makeatother
\begin{document}

\preprint{AIP/123-QED}

\title[Entangleware Sequencer]{Entangleware Sequencer:\\A  Control Platform for Atomic Physics Experiments}
\author{N. Kowalski}

\author{N. Fredman}%
 
\affiliation{ 
Department of Physics, University of Illinois Urbana-Champaign, Urbana, Illinois, 61801, USA
}%

\author{J. Zirbel}
 \homepage{https://entangleware.com/}
\affiliation{%
Entangleware, 113 Moiso Lane, Pleasant Hill, California, 94523, USA
}%

\author{B. DeMarco}
\affiliation{Department of Physics, University of Illinois Urbana-Champaign, Urbana, Illinois, 61801, USA}

\date{\today}

\begin{abstract}
Experimental quantum physics and computing platforms rely on sophisticated computer control and timing systems that must be deterministic. An exemplar is the sequence used to create a Bose-Einstein condensate at the University of Illinois, which involves 46,812 analog and digital transitions over 100 seconds with 20~ns timing precision and nanosecond timing drift. We present a control and sequencing platform, using industry-standard National Instruments hardware to generate the necessary digital and analog signals, that achieves this level of performance. The system uses a master 10~MHz reference clock that is conditioned to the Global Positioning Satellite constellation and leverages low-phase-noise clock distribution hardware for timing stability. A Python-based user front-end provides a flexible language to describe experimental procedures and easy-to-implement version control. A library of useful peripheral hardware that can be purchased as low-cost evaluation boards provides enhanced capabilities. We provide a GitHub repository containing example python sequences and libraries for peripheral devices as a resource for the community. 
\end{abstract}

\maketitle
\section{\label{sec:Intro}Introduction}
Cold atom and qubit control experiments use ``real-time" computer control systems with a high-level of reproducibility to manage experimental timing. These experiments operate in a shot-mode, where a single run of the experiment produces data, and parameters are changed between runs. Measurements are completed by repeatedly running the experiment and averaging over many shots. The experiment control platform must ensure that measurements only vary when parameters are purposefully changed and not because of drift in timing. Hence, the timing within a shot must be ``real-time" to preserve coherence and control, and the control platform must be able to determinstically manage a mix of analog and digital outputs and inputs, including cameras, photon detectors, and radiofrequency (RF) sources.


There are limited commercial and open-source options available. Commercial options include National Instruments (NI) LabVIEW sotware and NI hardware, M-Labs ARTIQ software and hardware \cite{Kulik2018}, and Quantum Machines OPX+ hardware and the QUA programming language \cite{QuantumMachine}. Open-source options include Cicero Word Generator \cite{Keshet2013} or the \textit{labscript} programming suite \cite{Starkey2013}. Research groups also buy hardware and create their own control platforms \cite{Sitaram2021, Reisenbauer2022, Hosak2018}, primarily using NI or ADwin hardware. Alternative hardware includes the Cypress programmable system on chip (PSoC) \cite{Sitaram2021} and RedPitaya STEMLab \cite{Reisenbauer2022}.

There are three key features that differ between these systems: the user interface, the approach to timing, and interfacing with peripheral hardware. The user interface can be either graphical, like that found in LabVIEW or Cicero, or scripted like that found in ARTIQ. For a fully functional platform like LabVIEW or Python, this is a personal preference. 

The system we report in this manuscript, called Entangleware, is a commercial software product utilizing industry-standard NI hardware. An advantage of using the NI ecosystem is that software interfaces are preserved across hardware generations. As we will discuss, Entangleware is a scripted system that supports relative and absolute timing and a rich ecosystem of interfacing with peripheral hardware via serial communication.

For any control system, the timing software must describe logically separate steps, possibly interleaved in time, with the ability to vary the relative timing between steps. There are different approaches to accomplishing this, including absolute and relative timing. A relative timing system is one in which each step occurs following the completion of the previous step. The steps in an absolute timing system occur at a time after the start time of the sequence.

As an example, we compare timing of sequence commands in Cicero with the system presented here. Both Cicero and Entangleware are compatible with some level of modular timing. Cicero divides time steps into ``words," where each word starts at the conclusion of the previous word (i.e., relative timing between steps), which is represented as columns in the graphical front end. This relative-timing only approach is limited in its ability to handle multiple steps happening simultaneously with different time scales. For example, consider a complex evaporative cooling stage that ramps multiple rf-frequency signals with differing trajectories and start times, but all interwoven in time. With only relative timing, this becomes quite complex, but it is conveniently described in the Entangleware scripted language using a combination of absolute and relative timing.

There are also two different approaches to timing on the hardware level: precompiled vs in-loop. In a precompiled system, such as Entangleware, the entire set of instructions is compiled and ready to run before the start of the actual experimental sequence. This allows the platform to support bitstreams of infinite length at the fastest timing possible across all channels. However, because the instructions are precompiled, there is only limited ability for in-loop decision making. An in-loop system compiles the instructions as the sequence runs. This supports complex decision making during a shot, but for long streams of instructions the sequence can out-pace the compilation.

Peripheral hardware interfacing also differs between platforms. Some types of hardware, such as USB devices, can not be used in a real-time system.  For Entangleware, we have developed a library of application programming interfaces (APIs) that make it easy to use  a suite of peripheral boards via serial programming interface (SPI) communication. Adding new boards to this library is straightforward. In contrast, ARTIQ includes a native AD9910 direct digital synthesizer (DDS) as a peripheral hardware board, but access to the board is restricted to the native functions in ARTIQ and does not offer full functionality. Simple commands like frequency sweeps are difficult to implement in ARTIQ, but easily obtainable with our system. Cicero has capabilities for RS-232 serial communication, but many peripherals are incompatible with this communication protocol. LabVIEW is also capable of SPI communication, like Entangleware, and contains many drivers for common hardware devices.

\section{\label{sec:Architecture}Entangleware Architecture}
\begin{figure}[h]
    \includegraphics[width=\linewidth]{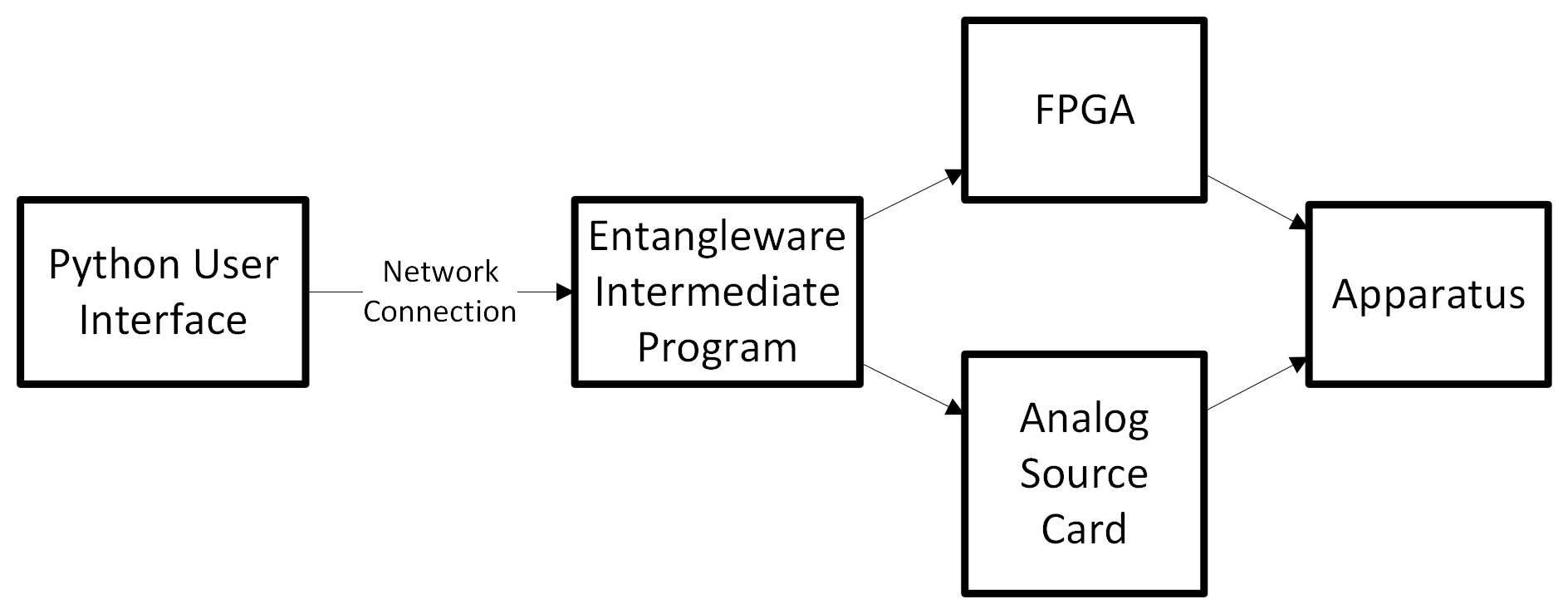}
    \caption{Entangleware system architecture.}
    \label{fig:toplevelflowchart}
\end{figure}

In the Entangleware architecture, a scripting language is used to describe a timed sequence of analog and digital events. This sequence is compiled into a bitstream that is sent to the Entangleware Control Application (ECA), which is an intermediate LabVIEW executable (Fig. \ref{fig:toplevelflowchart}). The compilation time is optimized for speed, taking 0.393 seconds for a neutral atom optical lattice sequence containing 129042 total transitions used in our group for measurements of diffusion in ultracold lattice gases. The ECA converts this bitstream into Verilog HDL and a set of instructions for the low-noise analog signal hardware. These sets of instructions are run once, and then the process is repeated with the sequence for the next shot. 

For Entangleware, any scripting language can be used to describe a sequence; we choose to use Python. A higher-level language allows for experimental parameters to be maintained in a human-readable format and facilitates logical separation of sequence blocks into modules. The user-written sequence needs to create at the lowest level a list of transitions with both the desired analog or digital state and the time at which the transition should occur. We divide the generation of these states into a hierarchy of sub-sequences, aiding legibility and facilitating relative timing as discussed above. Lower level sequences reflect the underlying hardware (e.g., sequences for magnetic coils), while mid-level sequences reflect the stage of the experiment (e.g., evaporative cooling).  Python is used to ``compile" these sub-sequences into a bitstream containing the transitions and times, which is sent to the ECA via a network interface. This has two advantages. First, it means the backend is agnostic to the user programming language, allowing for a high degree of user customization. Second, it is not necessary for the user control computer to be the same device as the intermediate hardware computer. The computer hosting the NI hardware can be placed next to the experimental apparatus and easily controlled from outside the room. 

The ECA manages the baseline FPGA and analog hardware source resources, as well as ``compiling" the timing information. The commands are translated into Verilog hardware description language and used to update the FPGA and analog source cards.

Currently Entangleware supports a NI PCIe-7820R FPGA for digital transitions and multiple analog source cards (see Table \ref{tab:Hardware}). The FPGA provides 128 digital input/output channels with 20ns timing. Some digital signals are used to directly control hardware. Most of the digital lines are used for serial control of peripheral hardware (e.g., AD9959 direct digital synthesizers acting as RF sources); see section \ref{sec:peripheral} for more information. Analog lines are used as references to servo lasers, microwave sources, and magnetic coils. Each NI PCI-6733 analog source card has eight 16-bit channels providing $\pm$10V ($\approx$0.3mV per step). 

\begin{table}
\caption{\label{tab:Hardware}Supported Control Hardware}
\begin{ruledtabular}
\begin{tabular}{ ll } 
 \multicolumn{2}{c}{Digital FPGA}\\
 \hline
 PCIe/PXIe-7820R & 128 Channels, 50 MHz clock\\ 
 \hline
 \multicolumn{2}{c}{Analog Sources}\\
 \hline
 PXIe/PCIe-6363 & 4 channels \\ 
 PCI/PXI-6733 & 8 channels \\
 PCIe/PXIe-6738 & 32 channels \\
 PXIe-6739 & 64 channels \\
\end{tabular}
\end{ruledtabular}
\end{table}

\section{\label{sec:Backend}Backend Architecture}
The ECA serves as the intermediary between a sequence description and the underlying hardware. In addition to describing the sequence, Python establishes communication with ECA. Upon launching the ECA software, it establishes a UDP listener on a known port. The Python code, when it launches, sets up a UDP messenger to query the network or host computer in search of the ECA. Additionally, Python configures a TCP server and relays the port number to the ECA through a UDP message. The ECA then establishes a connection with the Python TCP server, enabling the creation of a TCP writer and a TCP reader for efficient data communication between the ECA and Python.

Once the connection is established, Python can proceed to enqueue a sequence and transmit it to ECA. Several fundamental Python functions are used for interacting with the ECA. Two define the fundamental output commands, handling the states, times, and the specific digital or analog channel responsible for the transition. Other commands initiate and terminate the connection with the ECA, as well as instruct the ECA to commence or conclude queueing transition data. A comprehensive list of these basic functions can be found in Tables \ref{tab:ECACommand1} and \ref{tab:ECACommand2}.

The ECA receives the sequence bitstream from Python, which it organizes into chronological order, converts into a machine-readable format (i.e., Verilog), distinguishes between analog and digital commands, and triggers the hardware to begin the sequence. It subsequently manages the data stream to the hardware and ensures that the hardware remains synchronized throughout the experiment. Upon completion of the sequence, the ECA sends a message back to Python.

\begin{figure}[h]
    \centering
    \includegraphics[width=\linewidth]{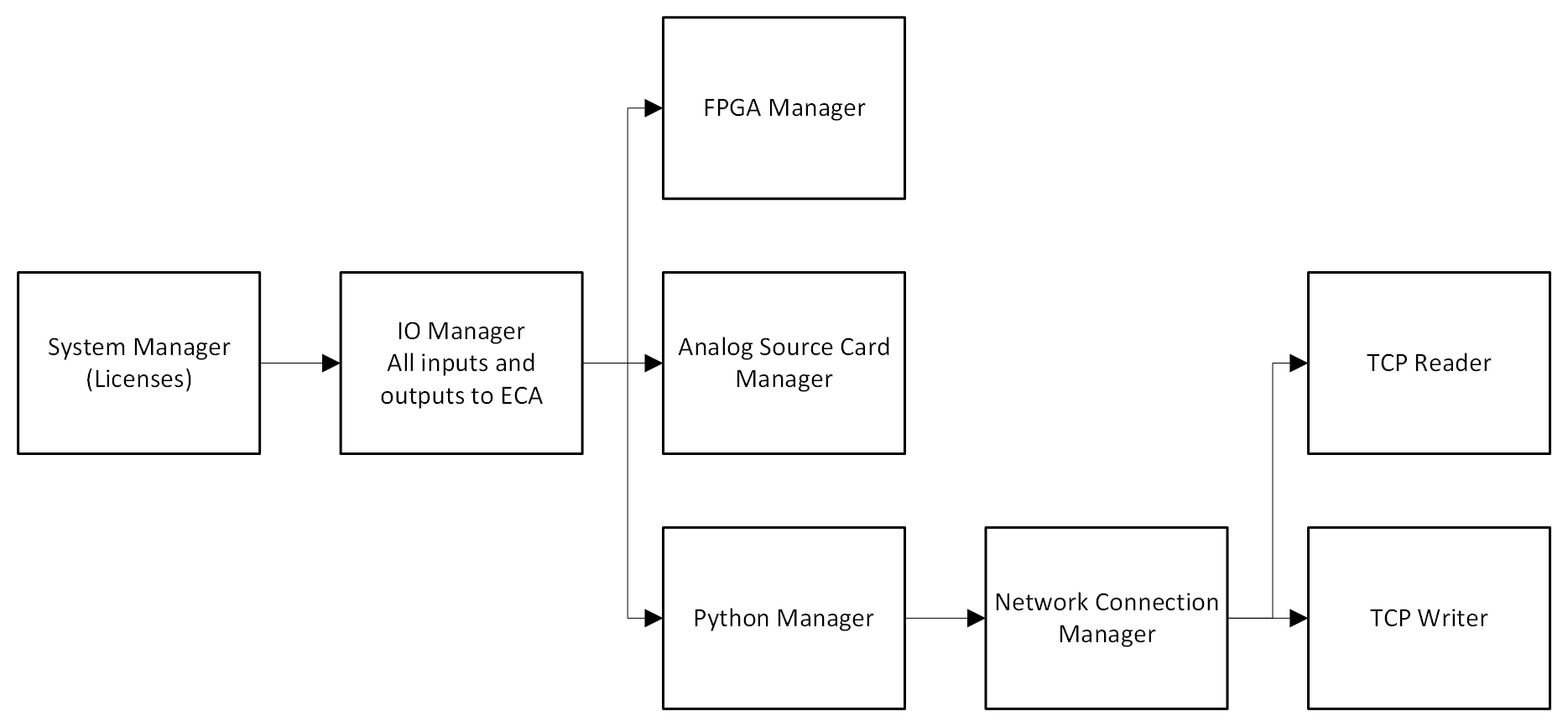}
    \caption{State machine object dependency structure of the ECA.}
    \label{fig:midlevelflowchart}
\end{figure}

\begin{table*}
\caption{\label{tab:ECACommand1}Table of Python run commands for the ECA.}
\begin{ruledtabular}
\begin{tabular}{p{0.11\linewidth} p{0.23\linewidth} p{0.56\linewidth}}
 Python Method & Arguments & Simple Description\\
 \hline
 connect & timeout\_sec: number of seconds to wait until timing out the TCP server & Connects to the ECA running on the local machine. This method should be called first, before any other methods listed here. \\
 disconnect & N/A & Disconnects from the ECA. Call when finished with the ECA, after a sequence is complete. \newline \\
 build\_sequence & N/A & Begins building a sequence which will queue deterministic commands for the ECA. \newline\\
 run\_sequence & N/A & Sends the queued sequence to ECA for further processing. This also tells the ECA to begin running the deterministic sequence. This method will return the number of queued elements in the sequence and the total time of the sequence in seconds. The method will block Python execution until the ECA has completed and will return a message. \newline \\
 run\_sequence\_chain & N/A & Similar to \textit{run\_sequence} except it will not block; allows another sequence to be queued while the ECA is busy executing the previous sequence. This helps to reduce dead-time. \newline \\
 clear\_sequence & N/A & Clears a sequence being queued. \newline \\
 rerun\_last\_sequence & N/A & Re-runs the previously executed sequence. This is useful to minimize communication latency with the ECA. \\
\end{tabular}
\end{ruledtabular}
\end{table*}

\begin{table*}
\caption{\label{tab:ECACommand2}Table of Python output commands for the ECA.}
\begin{ruledtabular}
\begin{tabular}{p{0.13\linewidth}p{0.15\linewidth} p{0.13\linewidth} p{0.15\linewidth} p{0.42\linewidth}}
 Python Method & \multicolumn{3}{c}{Arguments} & Simple Description\\
 \hline
\multirow{6}{*}{set\_digital\_state} & Name & Type (unit) & Description & \multirow{6}{0.9\linewidth}{\newline \newline \newline This sets the full state of digital channels/lines of a connector of the PCIe-7820 at a specified time. It uses the connector, the channels, the output enable state of the buffers, and the output state to define the full state. 
\begin{center}
\includegraphics[width=0.5\linewidth]{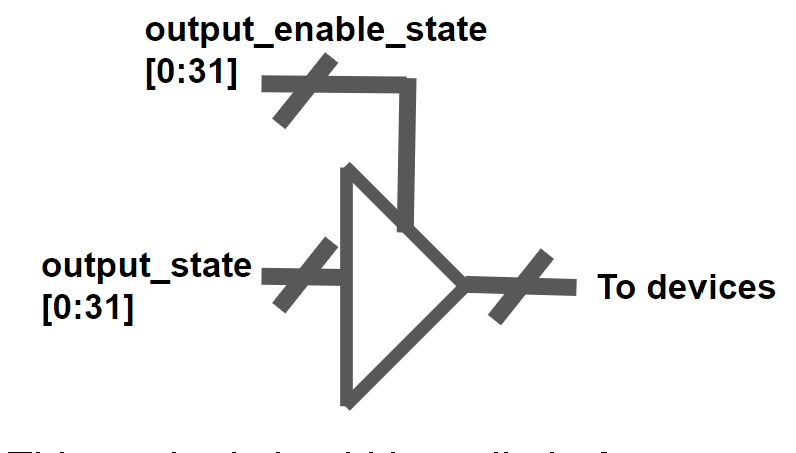} 
\end{center} 
This method should be called after \textit{build\_sequence} but before \textit{run\_sequence} to queue for deterministic sequences. Otherwise, this method ignores the time parameter and updates the PCIe-7820 immediately.}\\
\cline{2-4}
& time & double (seconds) & absolute time when the digital transition will occur \newline \\
& connector & U32 (N/A) & PCIe-7820 connector \newline \\
& channel\_mask & U32 (N/A) & bitwise mask of the digital channels to be set \newline \\
& output\_enable\_state & U32 (N/A) & bitwise mask which defines whether the output buffer is enabled (1) or not (0). Typically all bits will be 1. \newline\\
& output\_state & U32 (N/A) & bitwise mask which defines the output state of the output buffer of the selected line \\ \\ \\
\multirow{4}{*}{set\_analog\_state} & time & double (seconds) & absolute time when analog output will change on the PCI-6733 \newline & \multirow{4}{0.9\linewidth}{\newline \newline This method specifies the analog output voltage of a channel on a PCI-6733 board at a specific time.This method should be called after \textit{build\_sequence} but before \textit{run\_sequence} to queue for deterministic sequences. Otherwise, this method ignores the time parameter and updates the PCI-6733 immediately. If the time and value parameters are lists, then this method will queue the times and values with only a single method call.} \\
& board & integer (N/A) & board number 0 or 1 \newline \\
& channel & integer (N/A) & channel number between 0 and 7 on the PCI-6733 \newline \\
& value & double (volts) & voltage value between -10 and 10V \newline \\
\\
\end{tabular}
\end{ruledtabular}
\end{table*}

\subsection{\label{sec:subFPGA}FPGA}
The PCIe-7820 is responsible for the digital and analog timing for the system. All timing is derived from a 50 MHz oscillator which can be sourced externally via the EXTCLKIN pin of the PCIe-7820 Connector 0 (pin 67). If the external 50MHz oscillator is not present at ECA start-up, an internal 50 MHz oscillator is used instead. The PCIe-7820 supplies 128 digital outputs with LVTTL logic levels across the 4 output connectors (Connector [0:3]) and up to 8 digital outputs on the RTSI bus. For our system, the PCI-6733 analog cards share a single digital line from the RTSI bus as a primary clock. So, the PCIe-7820 is responsible for all timing of the system.

The full timing engine is a combination of 4+1 (connectors+RTSI) timing engines, each made from a counter and output latch. The counter is 56 bits wide, and the latch is 64 bits wide. Combined, they are responsible for the timing and output from a single connector. All counters/latches can be started synchronously using a shared on-board trigger. The reason 64 bits are needed for the latch is because, for each output line, two booleans are needed to represent its full state. One boolean represents whether the output buffer is enabled or tri-stated, the other boolean represents the output state. So for the 32 lines per connector, 64 bits are needed to fully represent the output state of the connector.  For our work, we only need outputs from the digital system, so the output buffer is always enabled. Future work may require using the DIO in a bidirectional manner for, e.g., in-loop decision making. 

Timing and output data are fed to each timing engine via a dedicated FIFO. At each tick of the 50MHz oscillator, the timing engine increments the count, compares it to the timing data, and decides whether to read and output new data or wait.

\begin{figure}[h]
    \centering
    \includegraphics[width=0.9\linewidth]{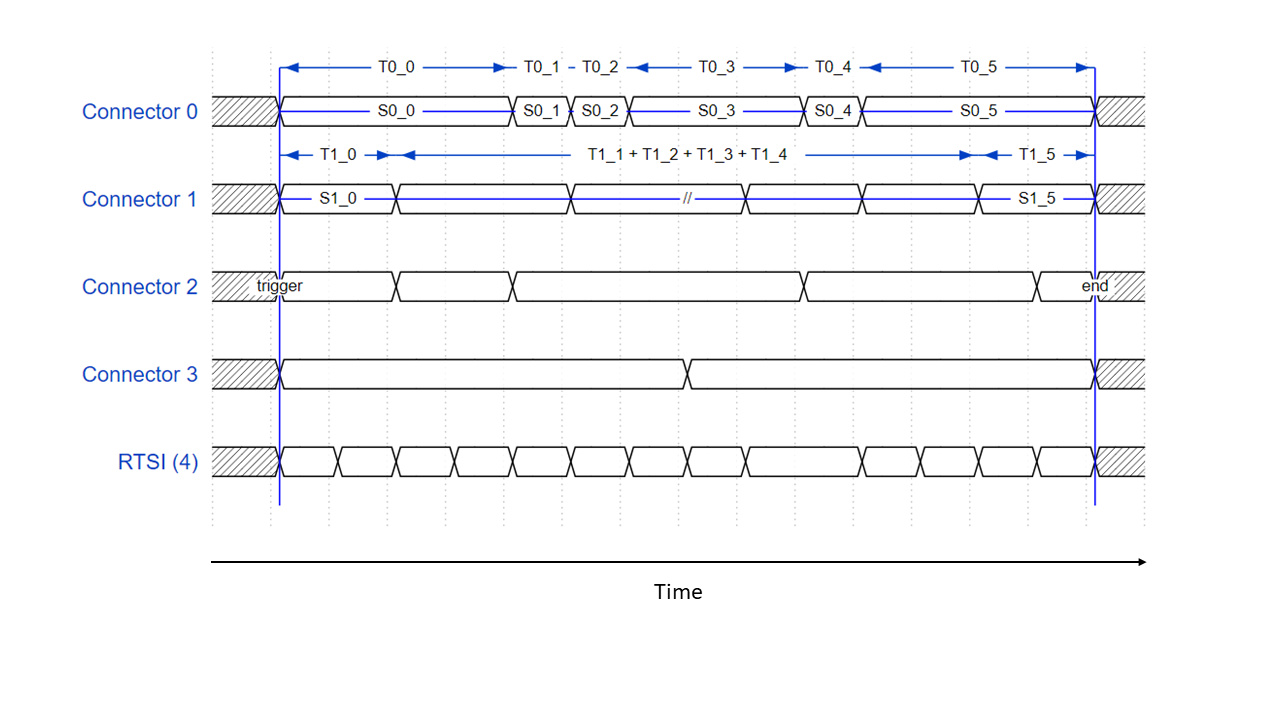}
    \caption{\textit{Example timing of all connectors and RTSI bus.}}
    \label{fig:FPGAtiming}
\end{figure}

Figure \ref{fig:FPGAtiming} shows an example timing diagram for all connectors/RTSI. The timing engine for connector $n$ is simply waiting for a trigger. Then, after the trigger, it transitions to state $Sn\_0$, and maintains that state for $Tn\_0$ ticks, and then transitions to state $Sn\_1$, and maintains that state for $Tn\_1$ ticks, etc.  The ECA is responsible for ensuring that all data is supplied to the timing engines and that all engines terminate at `end.'

Because the analog output from the PCI-6733 is limited to updating at a rate no faster than 0.769 MSPS across 8 channels, we chose a divisor of 80 for our 50~MHz base clock ($50/80 = 0.625$ MSPS).  This means the PCI-6733 timing ``grid" is 80 times longer than that of the PCIe-7820 timing grid. The ECA pads all of the digital timing data to ensure both the analog and digital grids terminate at the same time.

\begin{figure}[h]
    \centering
    \includegraphics[width=0.9\linewidth]{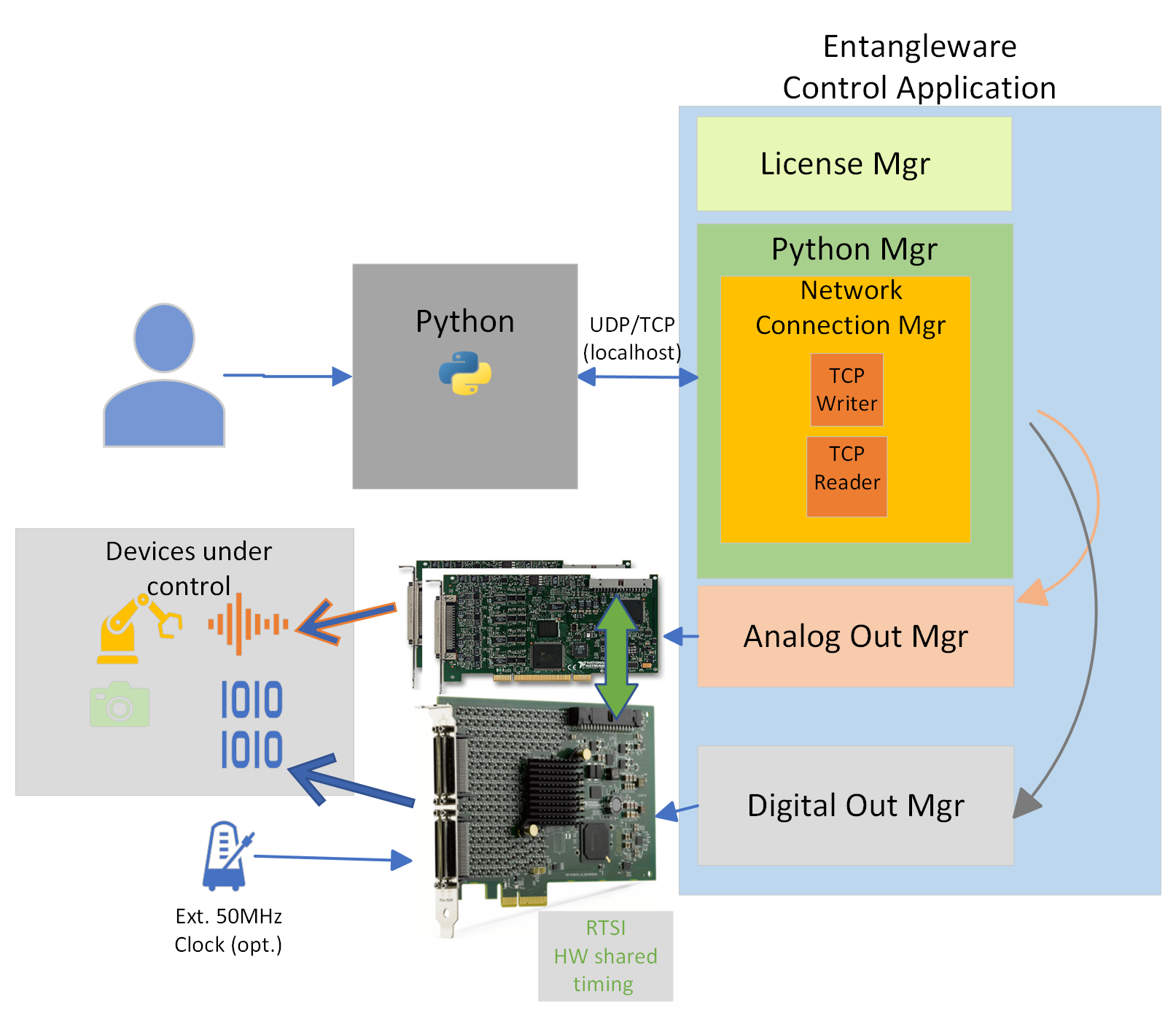}
    \caption{\textit{Structure of the ECA within the overall system architecture.}}
    \label{fig:ECAarchitecture}
\end{figure}

\subsection{\label{sec:subAnalog}Analog Cards}
Data is fed to the analog cards using the DAQmx interface from LabVIEW.  The clock for the cards is sourced by the FPGA via the RTSI bus.  The analog data is timed by the FPGA, but the analog values are sourced by the card itself.  The DAQmx API is responsible for controlling the configuration of the analog cards (e.g., reservation of analog card resources, specification of clock sources, starting and stopping DAQmx tasks, and disposing of the resources when finished). The analog cards possess a limited memory buffer. The remainder of the commands are stored in RAM on the control computer, and the DAQmx interface replenishes the buffer as analog commands are executed. 

\section{\label{sec:Clock}System Clocking}
We clock our experiment using a low phase-noise distribution and generation system to supply a frequency standard to many devices across a wide frequency range. As we will show in Section \ref{sec:Performance}, this results low drift and jitter for the digital, and hence analog, signals. A 10~MHz master clock is generated from a Stanford Research Systems PRS10 rubidum atomic clock that is conditioned to a one pulse-per-second GPS signal. This 10~MHz signal is the reference clock for all timing in the apparatus. A Stanford Research Systems FS710 distribution amplifier provides multiple copies of the 10~MHz signal, one of which is sent to an Analog Devices AD9510 clock distribution IC that provides eight different frequency clock outputs locked to the 10~MHz input by a phase-locked loop (PLL). This is how the 50~MHz clock signal for the FPGA is generated, as well as clock signals for all the RF sources in the experiment.

\begin{figure}[h]
    \centering
    \includegraphics[width=\linewidth]{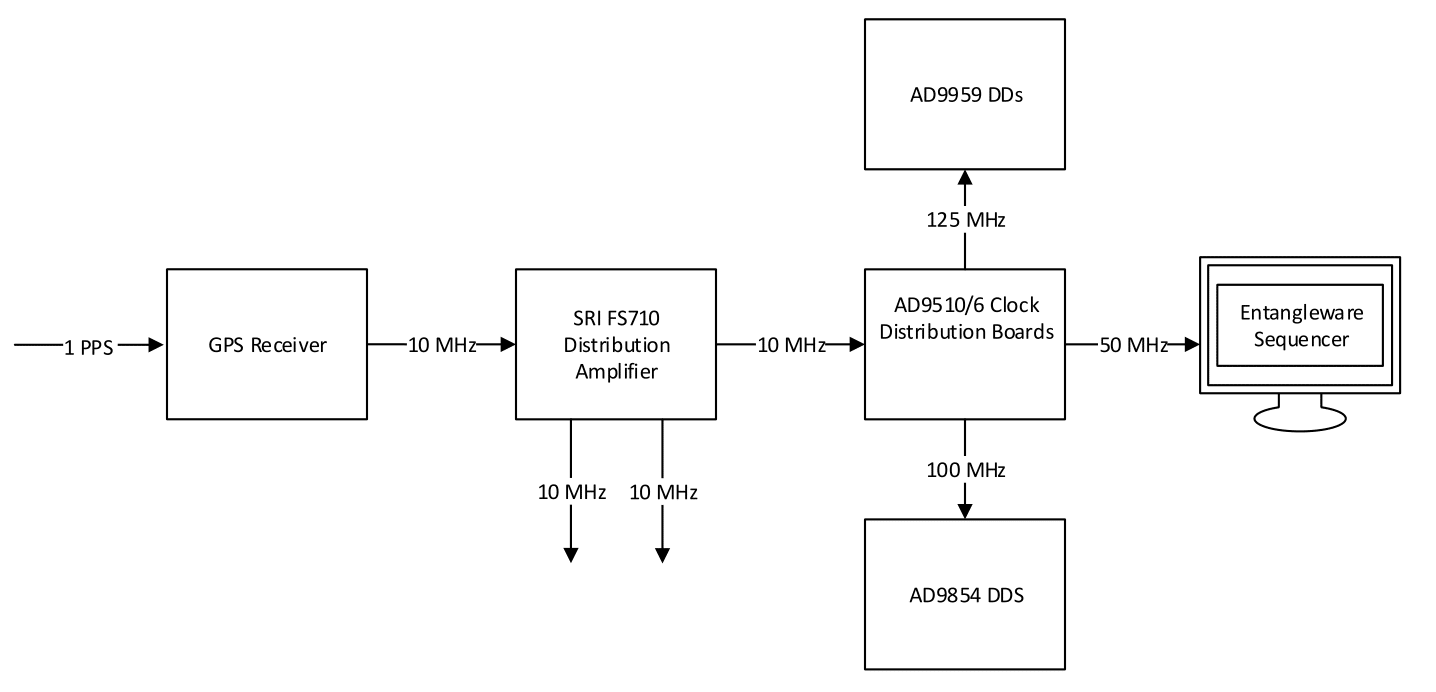}
    \caption{Timing network topology}
    \label{fig:Timing4}
\end{figure}

\section{\label{sec:peripheral}Peripherals}
We have developed the ability for this platform to interface with peripheral hardware via serial communication. We use low-cost evaluation boards for  functionality, including DDS boards for RF frequency sources and digital-analog-converters (DACs) for analog sources, as shown in Table \ref{tab:PeripheralTable}. The table provides the current cost of a fully functional evaluation board from Ananlog Electronics; the AD9854 evaluation board is no longer produced, although various third-party options are available. A library to control these boards is provided as a service to the community. This library of APIs allows the user to use simple commands to access the functionality of the peripheral hardware, while hiding the low-level serial communication, e.g. setting the output frequency of a DDS while hiding the code to generate and send the bitstream to the hardware. 

\begin{table}
\caption{\label{tab:PeripheralTable}Peripheral Hardware}
\begin{ruledtabular}
\begin{tabular}{ llc } 
 Board & Description & Cost \\ 
 \hline
 AD9959 & 4 Channel 500 MSPS DDS & \$558 \\ 
 AD9912 & 1 GSPS DDS & \$695 \\ 
 AD9910 & 1 GSPS DDS, arbitrary waveform generator & \$668 \\ 
 AD9854 & 300 MSPS Quadrature Complete DDS & N/A \\ 
 AD5372 & 32 Channel 16 bit DAC 540 KSPS & \$113 \\ 
\end{tabular}
\end{ruledtabular}
\end{table}

Each command (see Tables \ref{tab:CommandsAD5372}, \ref{tab:CommandsAD9854}, and \ref{tab:CommandsAD9959}) generates a digital bitstream that is sent to the memory buffer of the peripheral device. This ensures that instructions are sent to the device before the command is to be called, and then the device is triggered to enact the instructions at the appropriate time within the sequence. Consequently,  peripheral hardware outputs with deterministic timing is ensured, i.e. a change in DDS frequency happens at the correct time in the sequence with minimal jitter. 

Serial communication uses four digital lines: IO, serial clock, update, and reset. Communication with all of these boards is functionally similar. Python is used to generate a command bitstream to send to the board. The first byte of this bitstream is the address of the command register being addressed, and the remaining bytes are the desired commands within that register. An output function strobes the serial clock line to time communication between the sequencer and peripheral board, while the waveform is output on the IO line. The hardware stores the commands in an on-board memory buffer and executes them when the update line is pulsed on/off. 

\begin{table}
\caption{\label{tab:CommandsAD9959}AD9959 Commands}
\begin{ruledtabular}
\begin{tabular}{ lp{0.3\linewidth}p{0.4\linewidth} } 
Command & Arguments & Simple Description\\
 \hline
 \texttt{arb} & \texttt{start\_time, channel, frequency\_list, power\_list, n\_step, total\_time} & Receives a list of \texttt{n\_step} frequencies and powers to output. Starting at t=\texttt{start\_time} runs through the list, outputting each set uniformly spaced within time \texttt{total\_time} \newline \\
 \texttt{amplitude\_mod} & \texttt{start\_time, channel, a0, a1} & Varies the output amplitude of output channel \texttt{channel} between values \texttt{a0} and \texttt{a1} starting at time \texttt{start\_time} according to the voltage applied to an additional profile pin on the evaluation board. \newline \\
 \texttt{freq\_mod} & \texttt{start\_time, channel, f0, f1, ramptime, rampstep} & Varies the output frequency of channel \texttt{channel} between frequencies \texttt{f0} and \texttt{f1} in n=\texttt{rampstep} discrete steps at a rate determined by \texttt{ramptime} according to the voltage applied to a profile pin on the evaluation board.\\
 \end{tabular}
 \end{ruledtabular}
\end{table}

\begin{table}
\caption{\label{tab:CommandsAD9854}AD9854 Commands}
\begin{ruledtabular}
\begin{tabular}{ lp{0.3\linewidth}p{0.4\linewidth} } 
Command & Arguments & Simple Description\\
\hline
\texttt{arb} & \texttt{start\_time, chirp, frequency\_list, power\_list, n\_step, total\_time} &
Receives a list of \texttt{n\_step} frequencies and powers to output. Starting at t=\texttt{start\_time} runs through the list, outputting each set uniformly spaced within time \texttt{total\_time}. \texttt{chirp}enables chirp mode for wide-bandwidth frequency sweeps.\\
\end{tabular}
\end{ruledtabular}
\end{table}

\begin{table}
\caption{\label{tab:CommandsAD5372}AD5372 Commands}
\begin{ruledtabular}
\begin{tabular}{ lp{0.3\linewidth}p{0.4\linewidth} } 
Command & Arguments & Simple Description\\
\hline
\texttt{set} & \texttt{time, chan, value} & 
Calls output channel \texttt{chan} to output voltage \texttt{value} at time \texttt{time}\\
\end{tabular}
\end{ruledtabular}
\end{table}

In our experiments AD9959 DDSs are used as RF frequency sources for acousto-optic modulators, beat-note locks, and phase-locked loops for lasers and microwave-frequency sources. AD9854 DDSs are used as frequency sources for RF evaporation in a magnetic quadrupole trap, and AD9910 DDSs are used as additional RF frequency sources. The AD5372 DACs are used as additional analog sources for servos and feedforwards. 


\section{\label{sec:PythonUI}Python User-Interface}
The hardware has two fundamental outputs: analog transitions and digital transitions. In python we created two primitive functions
\begin{python}
analog_out(time, connector, channel, value)
digital_out(time, connector, channel, state)
\end{python}
that reflect the nature of the platform and are the basis of every sequence constructed. Each command receives as parameters the time at which the transition is to occur, the connector the output is found on (the FPGA has four VHDCI hardware output connectors, and there are two analog source cards), the line on the connector to be changed, and the value the line should change to. The steps of the sequence that utilize these primitive functions are organized using the class structure of python, creating a modular code base. 

A timing class (found in Appendix \ref{app:Timing}) handles absolute and relative timing for this functions and the higher level sequences that contain them. The timing class has two parameters, \texttt{start\_time} and \texttt{current\_time}, two methods \texttt{abs} (for absolute timing) and \texttt{rel} (for relative timing), and a decorator \texttt{\_update\_time}. Every sequence and sub-sequence is a daughter class of this \texttt{Sequence} class, and thus inherits these parameters and methods. \texttt{start\_time} is the start time of the sequence, and \texttt{current\_time} is incremented for each step within the sequence. Both methods take as an argument a command to be executed (e.g. a step like turning a laser on) and a time at which to execute that command. The method \texttt{abs} executes the command at time \texttt{t\_step} after \texttt{start\_time}. Method \texttt{rel} executes the command time \texttt{delay\_time} after the end of the previous command in the sequence. All sequences return the elapsed time at the end. The decorator \texttt{\_update\_time} is used for methods of sequence classes and automatically updates \texttt{start\_time} and \texttt{current\_time} to be the time at which the method is called. It also returns the elapsed time for the method. 

\begin{figure}[h]
    \centering
    \includegraphics[width=\linewidth]{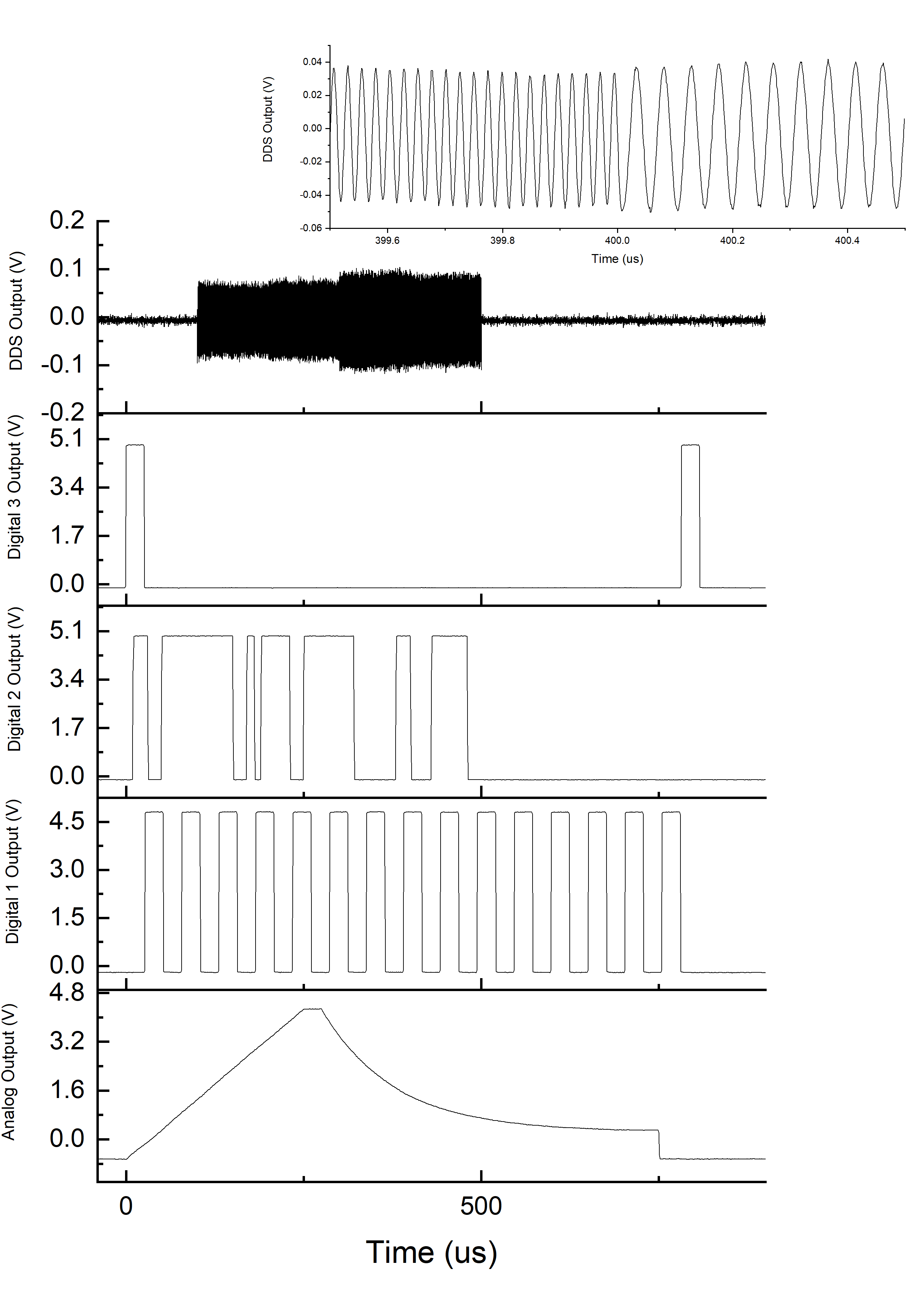}
    \caption{Measured analog, digital, and DDS outputs for \texttt{ExampleSequence}. The inset at the top shows a discrete transition from 40~MHz to 20~MHz output frequency for the DDS.}
    \label{fig:ExampleSequenceOutput}
\end{figure}

An example of a sequence that uses relative and absolute timing is presented in Appendix \ref{appendixExampleSequence}. This class, a daughter class of the \texttt{Sequence} class, ramps an analog output on and off, uses three different sets of digital pulses, and calls for a DDS to ramp an output frequency. The timing of these steps is set within the \texttt{seq} method, using the \texttt{abs} and \texttt{rel} methods from the parent \texttt{Sequence} class. The outputs of the relevant analog and digital lines and DDS for \texttt{ExampleSequence} are shown in figure \ref{fig:ExampleSequenceOutput}. 

To demonstrate the modularity of sequencing in Python the digital pulses, analog ramps, and DDS frequency sweep are defined in separate classes that are initialized and called within the \texttt{ExampleSequence} class, shown in Appendix \ref{app:PythonModule}. The analog ramp class procedurally generates a list of analog transitions that correspond to a desired ramp trajectory. Rather than call for transitions at a set rate (e.g., one transition per microsecond), we minimize the number of transitions by calculating when each DAC bitflip should occur for the desired ramp. Digital outputs are also procedurally generated in the \texttt{DigitalTransitions} class, another daughter of the \texttt{Sequence} class. The \texttt{DDSRamp} class passes a series of frequency tuning words (FTWs) to the DDS. By procedurally generating the FTWs, the DDS can perform a variety of frequency ramps. The phase-continuous frequency change when a new FTW is written can be seen in the inset of figure \ref{fig:ExampleSequenceOutput}.

\section{\label{sec:Performance}Timing Performance}

Timing stability is a key metric of system performance. To characterize the timing performance, we measured the relative timing between a digital line and the 10~MHz master clock (see section \ref{sec:Clock}). Since the digital signals set the analog timing, this measurement also characterizes the analog timing performance. We characterize the performance through two measurements. First, we repeatedly measure where a digital transition occurs relative to the master clock and characterize the jitter by the standard deviation of the difference. Second, we measure the drift in timing between the clock and the digital transition over ten seconds.

\begin{figure}[h]
    \centering
    \includegraphics[width=\linewidth]{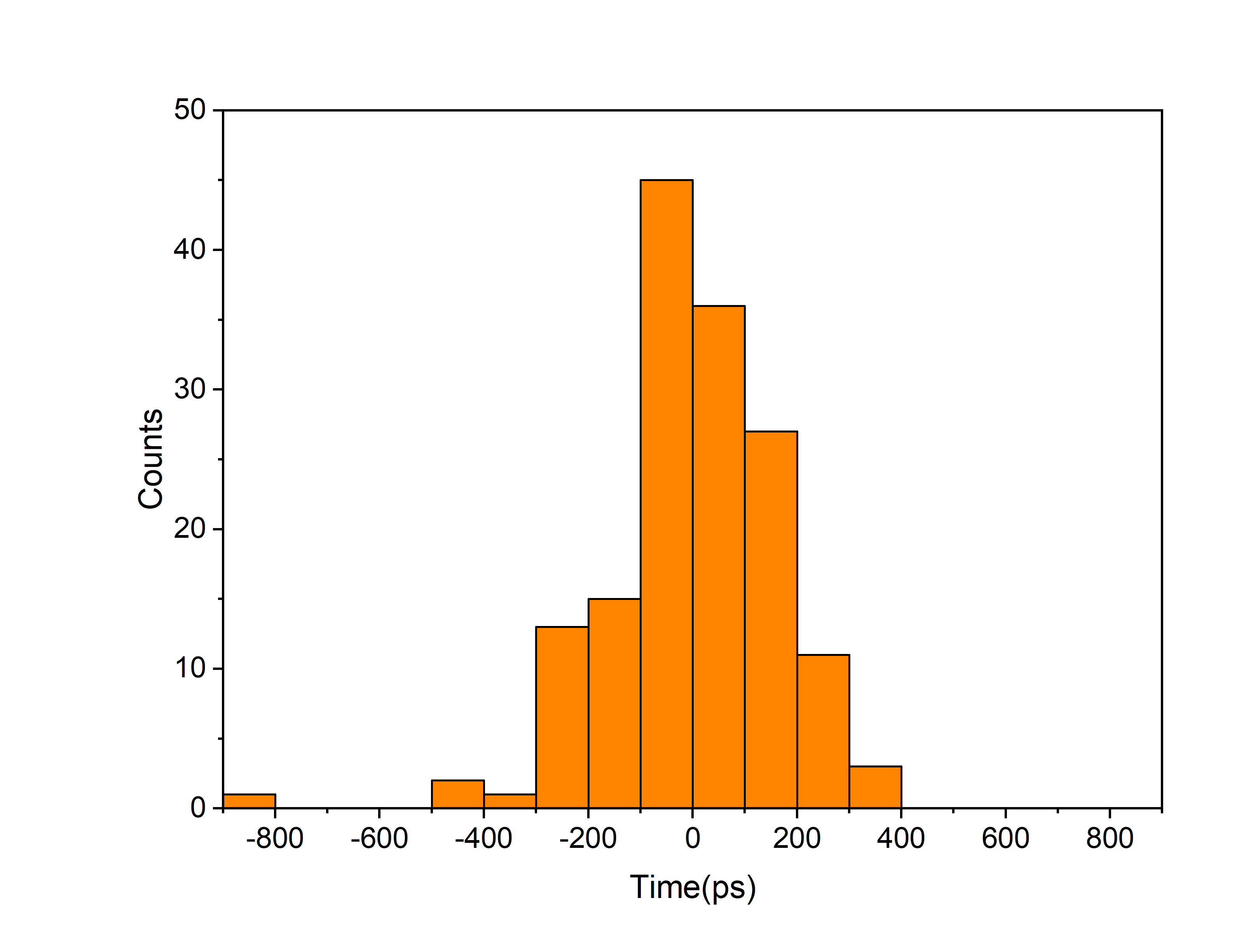}
    \caption{Time difference between a digital transition and the master clock signal. The distribution is constructed from 200 independent measurements.}
    \label{fig:TimingJitter}
\end{figure}

Figure \ref{fig:TimingJitter} shows the timing jitter between the 10~MHz clock signal and one of the digital outputs.  To consistently measure the timing of the clock we fit the signal to a sine wave. The time at which the digital signal passes a threshold voltage of 1.5 volts (chosen as approximately halfway between the TTL logic off and on values) is compared to the zero crossing of the fit.


We observed five distinct peaks relative to the 10~MHz master clock signal. This occurs because when the sequence is initiated, it will begin on the first 50~MHz clock zero-crossing. Our observation of five distinct peaks indicates a stable phase reference between the clocks. To obtain a single-peaked histogram, we subtract $2\pi/5$ from mean of the distributions to center them at $t=0$, and then convert from phase difference to time. The standard deviation of the distribution is $160$~ps, which is larger than the $50$~ps time resolution of the measurement. This variation has no effect on the system performance for our measurements, as the sub-nanosecond scale is many orders of magnitude faster compared with signaling times. 

\begin{figure}[h]
    \centering
    \includegraphics[width=\linewidth]{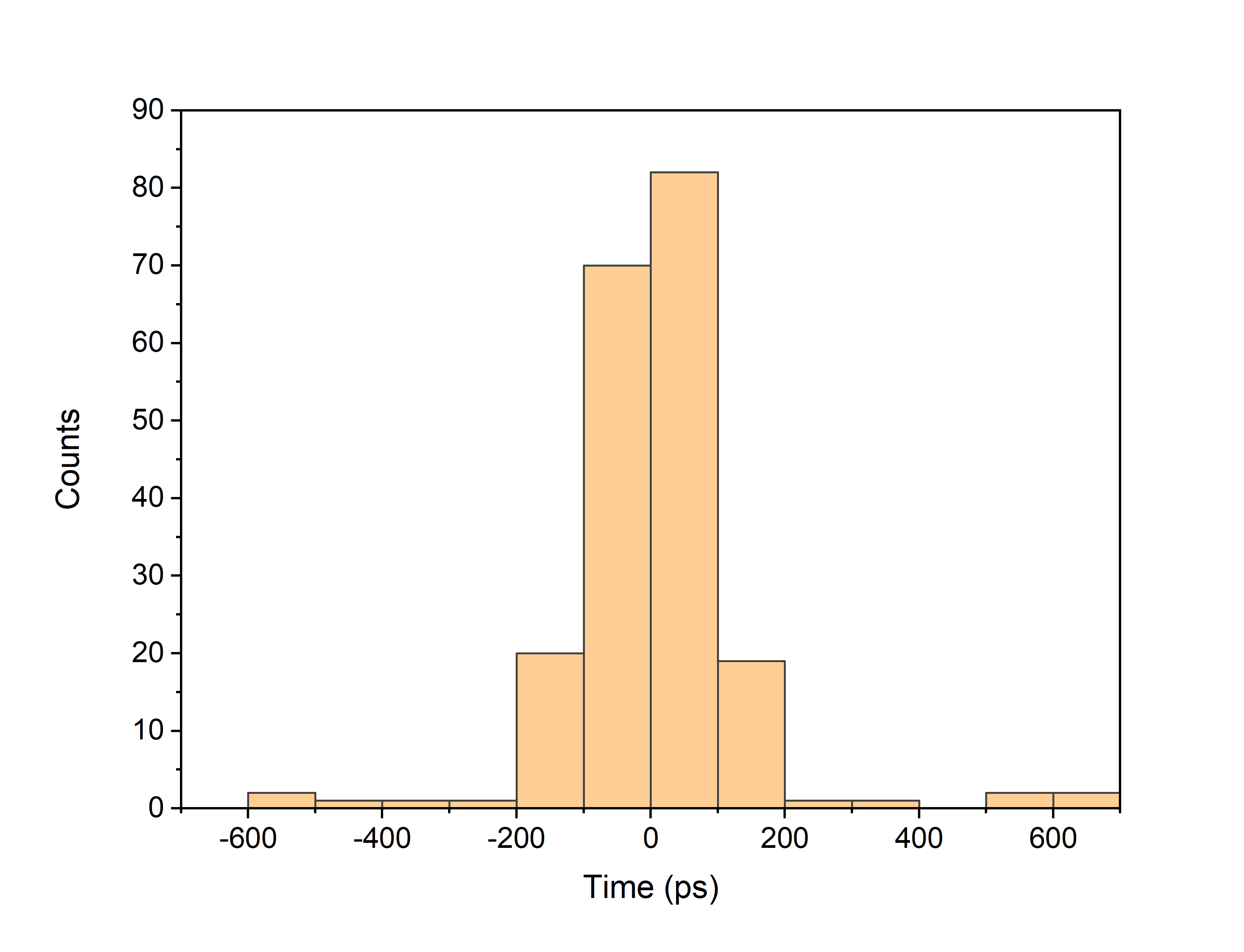}
    \caption{Time drift between the 10~MHz clock and a digital output over the course of a 10 second sequence. This distribution was constructed from 120 independent measurements.}
    \label{fig:CoherenceDataSubtracted}
\end{figure}

To measure there long-term drift, we repeated the this measurement in the same shot at  $t=0$ and $t=10$~s. For each run we, subtracted the phase difference at the two time, producing the distribution shown in Fig. \ref{fig:CoherenceDataSubtracted}. The mean of this distribution is $2$~ps, indicating negligible drift.

\section{Conclusion}

In conclusion, we have described a real-time control system that can be used for atomic physics experiments. It relies on standard National Instruments hardware, the commercially available software package Engtangleware, and commercially available timing hardware. The system can produce digital and analog signals with minimal timing jitter and drift. A library of  peripheral drivers and example code can be found at \url{https://github.com/DeMarcoAMO/Entangleware}.

\section{Acknowledgments}

We acknowledge support from NSF grant PHY-2110291.

\clearpage

\appendix

\section{\label{app:Timing}Timing Class}

In this appendix, we provide the base timing classes that are used to handle absolute and relative timing within sequences. 

\begin{widetext}
\begin{python}
# class to handle timing within sequences
class Sequence:
    def __init__(self):
        self.start_time = 0
        self.current_time = self.start_time

    # executes sequence seq at time t_step
    def abs(self, t_step, seq=null_func):
        self.current_time = t_step + self.start_time
        step_time = seq(self.current_time)
        self.current_time += step_time
        return step_time

    # executes seq at time delay_time after the previous sequence is finished executing
    def rel(self, delay_time, step=null_func):
        self.current_time += delay_time
        if type(step) is list:
            for seq in step:
                step_time = seq(self.current_time)
            self.current_time += step_time
            return step_time
        else:
            step_time = step(self.current_time)
            self.current_time += step_time
            return step_time

    # decorator to update the values of absolute time and current time in a daughter class
    def _update_time(func):
        def time_wrapper(self, t):
            self.start_time = t
            self.current_time = t
            func(self, t)
            time_elapsed = self.current_time - self.start_time
            return time_elapsed
        return time_wrapper
\end{python}
\end{widetext}

\section{\label{appendixExampleSequence}Example Sequence}

Here, we provide an example sequence that calls a series of digital pulses, analog ramps, and a DDS frequency ramp using the timing methods found in Appendix \ref{app:Timing}. The pulses and ramps are generated using python classes found in Appendix \ref{app:PythonModule}. The output of this sequence is shown in Fig. \ref{fig:ExampleSequenceOutput}.

\begin{widetext}
\begin{python}
class ExampleSequence(Sequence):
    def __init__(self):
        super().__init__()

        # initialize 2 analog ramp classes, one for a linear up ramp and the other exponential down
        self.ramp_up = AnalogRamp(v_start=0, v_end=5, ramp_time=200*us)
        self.ramp_down = AnalogRamp(v_start=5, v_end=1, ramp_time=400*us, tau=100*us)
        self.analog_off = partial(out.analog_out, connector=1, channel=3, value=0)

        # initialize digital line class
        self.digital = DigitalTransitions()

        # initialize DDS sweep class
        self.dds = DDSRamp(freq_start=80*MHz, freq_stop=5*MHz, ramp_time=1*ms, n_steps=25)

    @Sequence._update_time
    def seq(self, start_time):
        self.abs(0.00, self.ramp_up.linear)
        self.rel(100*us, self.ramp_down.exponential_down)
        self.rel(50*us, self.analog_off)

        self.abs(250*us + 270*ns + 880.15*us, self.digital.pulse3)
        self.rel(0.00, self.digital.pulse1)
        self.rel(0.00, self.digital.pulse3)
        self.abs(0.5*ms, self.digital.pulse2)

        self.abs(250*us, self.dds.linear)
\end{python}
\end{widetext}

\section{\label{app:PythonModule}Python Modules}

Here we document the lower-level sequences used within \texttt{ExampleSequence}: \texttt{AnalogRamp}, \texttt{DigitalTransitions}, and \texttt{DDSRamp}. These sequences demonstrate the modularity of our approach and the ability to programatically generate commands. \texttt{ExampleSequence} interweaves these commands in time, as seen in figure \ref{fig:ExampleSequenceOutput}.

\subsection{\label{app:subAnalogRamp}Analog Ramp}
\texttt{AnalogRamp} addresses channel 3 on PCI-6733 board 1. Instances of the \texttt{AnalogRamp} class are initialized with a start voltage, and voltage, ramp time, and time constant (for an exponential curve). The start and end voltages are converted into bit values for the analog source card, and a list of all the intermediary bit values is generated. The methods linear and exponential calculate the corresponding times at which each bit value should be output according to the ramp time (and time constant), then output the list of voltages and times using the \texttt{\_output} method. 
\begin{widetext}
\begin{python}
# create a class to handle continuous analog ramps
# Figure out when each bit flip should occur for the DAC
class AnalogRamp:
    # ramp from voltage v_start to v_end over time ramp_time
    # if exponential use time constant tau
    def __init__(self, v_start, v_end, ramp_time, tau=0):
        super().__init__()
        self.tau = tau
        self.ramp_time = ramp_time
        # output to channel 3 on board 1:
        self.board = 1
        self.channel = 3
        # convert the start/end values into their corresponding DAC bits
        self.q_val_start = int((v_start / 20) * (2 ** 16))
        self.q_val_end = int((v_end / 20) * (2 ** 16))
        # generate a list of DAC bits between q start and q end
        if self.q_val_start > self.q_val_end + 1:
            self.output_steps = list(range(self.q_val_end + 1, self.q_val_start))
        else:
            self.output_steps = list(range(self.q_val_start, self.q_val_end + 1))
        self.length = len(self.output_steps)
        # generate blank lists for the actual analog steps and their corresponding times
        self.time_steps = [None] * self.length
        self.analog_steps = [None] * self.length

    # iterates through the list of analog steps outputting each one at the appropriate time
    def _output(self):
        for index in range(self.length):
            time = self.time_steps[index]
            value = self.analog_steps[index]
            out.analog_out(time, self.board, self.channel, value)

    # populates output lists for a linear ramp then calls output method
    def linear(self, t_start):
        # slope of linear ramp
        slope = (self.q_val_end - self.q_val_start) / self.ramp_time
        # calculate the times and convert the outputs back into voltages
        for index in range(self.length):
            self.time_steps[index] = t_start + (self.output_steps[index] - self.q_val_start) / slope
            self.analog_steps[index] = 20 * self.output_steps[index] / (2 ** 16)
        if self.q_val_start > self.q_val_end + 1:
            self.time_steps.reverse()
            self.analog_steps.reverse()
        self._output()
        return self.ramp_time

    # populates output lists for exponentially decreasing ramp then calls output
    def exponential_down(self, t_start):
        delta = math.exp(self.ramp_time / self.tau)
        if delta == 1:
            raise ValueError('decay_rate or the time interval is too small')
        alpha = (delta-1)/(self.q_val_start - self.q_val_end)
        for index in range(self.length):
            self.time_steps[index] = t_start + self.ramp_time - \
                                     self.tau*math.log(1 + \
                                     (self.output_steps[index] - \
                                     self.q_val_end)*alpha)
            self.analog_steps[index] = 20 * self.output_steps[index] / (2 ** 16)
        if self.q_val_start > self.q_val_end + 1:
            self.time_steps.reverse()
            self.analog_steps.reverse()
        self._output()
        return self.ramp_time
\end{python}
\end{widetext}

\subsection{\label{app:subDigital}Digital Transitions}
Here, we show the \texttt{DigitalTransitions} class that outputs to three different digital lines, channels 16, 18, and 20 on connector 2. The \texttt{partial} method is used to create six new functions, turning each digital line on and off. \texttt{partial} allows for some of the parameters of a function to be assigned, in this case the connector, channel, and state of \texttt{digital\_out}, leaving just the time parameter to be handled by the \texttt{self.abs} and \texttt{self.rel} methods of the parent \texttt{Sequence} class.

\begin{widetext}
\begin{python}
# create a sequence class to handle programatic generation of digital transitions
class DigitalTransitions(Sequence):
    def __init__(self):
        # initialize parent Sequence class
        super().__init__()
        # output on digital lines 16, 18 and 20 on VHDCI connector 2 of the FPGA
        self.connector = 2
        self.channel1 = 18
        self.channel2 = 16
        self.channel3 = 20

    # method to repeatedly pulse channel 1 starting at t=start_time
    @Sequence._update_time
    def pulse1(self, start_time):
        # pulse on and off in 1ms intervals 25 times
        pulse_time = 50*us
        n_pulses = 15
        # use partial method to fill the connector, channel, and state parameters of out.digital_out
        # creates functions on(time) and off(time) that accept one time parameter
        on = partial(out.digital_out, connector=self.connector, channel=self.channel1, state=1)
        off = partial(out.digital_out, connector=self.connector, channel=self.channel1, state=0)

        # pulse line on at t = 0 (start_time)
        self.abs(0.00, on)
        # pulse line off and on
        for n in range(n_pulses-1):
            self.rel(pulse_time, off)
            self.rel(pulse_time, on)
        self.rel(pulse_time, off)

    @Sequence._update_time
    def pulse2(self, start_time):
        # generate a list of pulse start times and lengths
        time_list = [0.01, 0.05, 0.17, 0.19, 0.25, 0.38, 0.43]
        pulse_length_list = [0.02, 0.10, 0.01, 0.04, 0.07, 0.02, 0.05]
        if len(time_list) != len(pulse_length_list):
            raise ValueError('Lists are of unequal lengths')

        # on/off functions for channel 2
        on = partial(out.digital_out, connector=self.connector, channel=self.channel2, state=1)
        off = partial(out.digital_out, connector=self.connector, channel=self.channel2, state=0)

        # iterate through list
        for n in range(len(time_list)):
            self.abs(time_list[n]*ms, on)
            self.rel(pulse_length_list[n]*ms, off)

    @Sequence._update_time
    def pulse3(self, start_time):
        self.abs(0.00, partial(out.digital_out, connector=self.connector, channel=self.channel3, state=1))
        self.rel(50*us, partial(out.digital_out, connector=self.connector, channel=self.channel3, state=0))
\end{python}
\end{widetext}

\subsection{\label{app:subDDSRamp}DDS Frequency Ramp}
In this section, we show \texttt{DDSRamp}, which uses an AD9854 DDS ramp to output an RF frequency that slews from \texttt{freq\_start} to \texttt{freq\_stop} in time \texttt{ramp\_time} in $n$ discrete steps. This is accomplished by programatically generating a list of frequency tuning words (FTWs) according to the desired ramp trajectory, and pushing them to the DDS at the appropriate time. Here we demonstrate a simple linear sweep, but more complex trajectories can easily be created. 
\begin{widetext}
\begin{python}
class DDSRamp(Sequence):
    def __init__(self, freq_start, freq_stop, ramp_time, n_steps):
        super().__init__()
        self.f_start = freq_start
        self.f_stop = freq_stop
        self.ramp_time = ramp_time
        self.n_steps = n_steps
        self.power = -18*dBm
        # initialize a AD9854 DDS board
        self.dds = brd.AD9854(connector=1, mosipin=29, sclkpin=25, resetpin=27, ioupdatepin=31, 
                              refclock=50e6, rmprateclk=24, f_ini=freq_start)
        # turn AD9854 off
        self.dds_off = partial(self.dds.arb, chirp=0, total_time=1*ms, freq=[0*MHz], 
                               power=[float('-inf')], num_steps=1)

    @Sequence._update_time
    def linear(self, start_time):
        # calculate FTWs for linear sweep from f_start to f_stop in t=ramp_time using n steps
        t_step = self.ramp_time/self.n_steps
        slope = (self.f_stop - self.f_start)/self.ramp_time
        freq_list = [self.f_start + slope*t_step*n for n in range(self.n_steps+1)]
        power_list = [self.power for n in range(self.n_steps+1)]
        # set up frequency sweep using calculated tuning words
        dds_sweep = partial(self.dds.arb, chirp=False, total_time=self.ramp_time, freq=freq_list, 
                            power=power_list, num_steps=self.n_steps)

        # initialize dds, do frequency sweep, and turn off
        self.abs(-10 * ms, self.dds.ini)
        self.abs(0.00, dds_sweep)
        self.rel(0.00, self.dds_off)
\end{python}
\end{widetext}

\subsection{\label{app:subRunScript}Run Script}
In this section, we show the Python commands used to run the \texttt{seq} method of \texttt{ExampleSequence} class. This code connects to the ECA, initializes and builds the bitstream, sends it to the ECA, and disconnects when the seqeuence is complete. 

\begin{widetext}
\begin{python}
# connect to ECA
ew.connect(1.0)

# states here are 'default' values
# set default of the analog and digital lines used to 0
out.digital_out(0.00, 2, 16, 0)
out.digital_out(0.00, 2, 18, 0)
out.digital_out(0.00, 2, 20, 0)
out.analog_out(0.00, 1, 3, 0)

# initialize bitstream to send to hardware
# everything between build_sequence and run_sequence is 
# deterministically timed
ew.build_sequence()

example = ExampleSequence()
example.seq(0.00)
# send bitstream to ECA
ew.run_sequence()

# disconnect from ECA
ew.disconnect()
\end{python}
\end{widetext}

\bibliography{bibliography}

\end{document}